\documentclass[prd,aps,showpacs,epsf,floats,onecolumn]{revtex4}
\usepackage{amsfonts}
\usepackage{amsmath}

\input{tcilatex}
\begin{document}

\title{REEXAMINATION OF AN INFORMATION GEOMETRIC CONSTRUCTION OF ENTROPIC
INDICATORS OF COMPLEXITY}
\author{C. Cafaro$^{1}$, A. Giffin$^{2}$, S. A. Ali$^{3}$, D.-H. Kim$^{4}$}
\affiliation{$^{1}$Dipartimento di Fisica, Universit\`{a} di Camerino, I-62032 Camerino,
Italy\\
$^{2}$Princeton Institute for the Science and Technology of Materials,
Princeton University, Princeton, NJ 08540, USA\\
$^{3}$Department of Physics, State University of New York at Albany, 1400
Washington Avenue, Albany, NY 12222, USA\\
$^{4}$Center for Quantum Spacetime, Sogang University, Shinsu-dong 1,
Mapo-gu, Seoul 121-742, South Korea }

\begin{abstract}
Information geometry and inductive inference methods can be used to model
dynamical systems in terms of their probabilistic description on curved
statistical manifolds.

\ In this article, we present a formal conceptual reexamination of the
information geometric construction of entropic indicators of complexity for
statistical models. Specifically, we present conceptual advances in the
interpretation of the information geometric entropy (IGE), a statistical
indicator of temporal complexity (chaoticity) defined on curved statistical
manifolds underlying the probabilistic dynamics of physical systems.
\end{abstract}

\maketitle

\section{Introduction}

The mystery of the origin of life and the unfolding of its evolution is
perhaps the most fascinating topic that motivates the description and, to a
certain extent, the understanding of the extremely elusive concept of
complexity \cite{L88, GM95, F98}. From a more pragmatic point of view, its
description and understanding is also motivated by the question of how
complex is quantum motion. This issue is of primary importance in quantum
information science, having deep connections to entanglement and
decoherence. However, our knowledge of the relations between complexity,
dynamical stability, and chaoticity in a fully quantum domain is not
satisfactory \cite{O98, C09}. The concept of complexity is very difficult to
define and its origin is not fully understood \cite{W84, W85, R95}. It is
mainly for these reasons that several quantitative measures of complexity
have appeared in the scientific literature \cite{GM95, L88}. In classical
physics, measures of complexity are understood in a better\textbf{\ }%
satisfactory manner. The Kolmogorov-Sinai metric entropy \cite{K65, K68},
the sum of all positive Lyapunov exponents \cite{P77}, is a powerful
indicator of unpredictability in classical systems. It measures the
algorithmic complexity of classical trajectories \cite{B83, B00, S89, W78}.
Other known measures of complexity are the logical depth \cite{B90}, the
thermodynamic depth \cite{S88}, the computational complexity \cite{P94} and
stochastic complexity \cite{R86} to name a few. For instance the logical and
thermodynamic depths consider complex (roughly speaking) whatever can be
reached only through a difficult path. Each one of these complexity measures
captures to some degree our intuitive ideas about the meaning of complexity.
Some of them just apply to computational tasks and unfortunately, only very
few of them may be generalized so that their applications can be extended to
actual physical processes. Ideally, a good definition of complexity should
be mathematically rigorous as well as intuitive so as to allow for the
analysis of complexity-related problems in computation theory and
statistical physics. For obvious reasons, a quantitative measure of
complexity is genuinely useful if its range of applicability is not limited
to a few unrealistic applications. For similar reason, in order to properly
define measures of complexity, the reasons for defining such a measure
should be clearly stated as well as what feature the measure is intended to
capture.

One of the major goals of physics is modeling and predicting natural
phenomena by using relevant information about the system of interest. Taking
this statement seriously, it is reasonable to expect that the laws of
physics should reflect the methods for manipulating information. Indeed, the
less controversial opposite point of view may be considered where the laws
of physics are used to manipulate information. This is exactly the point of
view adopted in quantum information science where information is manipulated
using the laws of quantum mechanics \cite{nielsen}. An alternative viewpoint
may be explored where laws of physics are nothing but rules of inference 
\cite{caticha-inference}. In this view the laws of physics are not laws of
nature but merely reflect the rules we follow when processing the
information that happens to be relevant to the physical problem under
consideration.

Inference is the process of drawing conclusions from available information.
When the information available is sufficient to make unequivocal, unique
assessments of truth, we speak of making deductions: on the basis of this or
that information we deduce that a certain proposition is true. In cases
where we do not have statements that lead to unequivocal conclusions, we
speak of using inductive reasoning and the system for this reasoning is
probability theory \cite{cox}. The word "induction" refers to the process of
using limited information about a few special cases to draw conclusions
about more general situations. Following this alternative line of thought,
we extended the applicability of information geometric techniques \cite%
{amari} and inductive inference methods \cite{c0, c1, c2, c3, c4} to
computational problems of interest in classical and quantum physics,
especially with regard to complexity characterization of dynamical systems
in terms of their probabilistic description on curved statistical manifolds.
Moreover, we seek to identify relevant measures of chaoticity of such an
information geometrodynamical approach to chaos (IGAC) \cite{carlo-MPLB,
carlo-PA, carlo-IJTP, carlo-PD, carlo-EJTP, carlo-CSF, carlo-PhD, carluccio}.

In this article, we present a formal and conceptual reexamination of the
information geometric entropy (IGE) \cite{carlo-PhD}, a statistical
indicator of temporal complexity (chaoticity) of dynamical systems in terms
of their probabilistic description using information geometry and inductive
inference.

We emphasize we do not present here any new application of the IGAC (for
instance, one of our most recent applications appears in \cite{carluccio}),
however (and, most importantly) we do report some relevant conceptual
advances in the interpretation of the IGE as a useful measure of complexity
for statistical models suitable for probabilistic descriptions of dynamical
systems.

The layout of this article is as follows. In Section II, we briefly review
our information geometric approach to the description of complex systems by
using information geometry and inductive inference. In Section III, we focus
on the key-steps leading to the construction of the IGE and on its
conceptual interpretation. Finally, in Section IV we present our final
remarks.

\section{Complexity on Curved Manifolds}

IGAC \cite{carlo-MPLB, carlo-PA, carlo-IJTP, carlo-PD, carlo-EJTP, carlo-CSF}
is a theoretical framework developed to study chaos in informational
geodesic flows describing physical systems. The reformulation of dynamics in
terms of a geodesic problem allows for the application of a wide range of
well-known geometric techniques to the investigation of the solution space
and properties of the equations of motion. All dynamical information is
collected into a single geometric object (namely, the manifold on which
geodesic flow is induced) in which all the available manifest symmetries of
the system are retained. For instance, integrability of the system is
connected with existence of Killing vectors and tensors on this manifold%
\textbf{.} The sensitive dependence of trajectories on initial conditions,
which is a key ingredient of chaos, can be investigated\textbf{\ }by using
the equation of geodesic deviation. IGAC is the information-geometric
analogue of conventional geometrodynamical approaches \cite{casetti, di bari}
where the classical configuration space\textbf{\ }$\Gamma _{E}$\textbf{\ }is
replaced by a statistical manifold\textbf{\ }$\mathcal{M}_{S}$\textbf{\ }%
with the additional possibility of considering chaotic dynamics arising from
non conformally flat metrics (the Jacobi metric is always conformally flat).
It is an information-geometric extension of the Jacobi geometrodynamics (the
geometrization of a Hamiltonian system by transforming it to a geodesic flow 
\cite{jacobi}). In the Riemannian \cite{casetti} and Finslerian \cite{di
bari} (a Finsler metric is obtained from a Riemannian metric by relaxing the
requirement that the metric be quadratic on each tangent space)
geometrodynamical approach to chaos in classical Hamiltonian systems, an
active field of research concerns the possibility\textbf{\ }of finding a
rigorous relation among the sectional curvature, the Lyapunov exponents, and
the Kolmogorov-Sinai dynamical entropy (i.e., the sum of positive Lyapunov
exponents) \cite{kawabe}.

An $n$-dimensional $%
\mathbb{C}
^{\infty }$ differentiable manifold (or more simply, a manifold) is a set of
points $\mathcal{M}$ admitting coordinate systems $\mathcal{C}_{\mathcal{M}}$
and satisfies the following two conditions: 1) each element $c\in \mathcal{C}%
_{\mathcal{M}}$ is a one-to-one mapping from $\mathcal{M}$ to some open
subset of $%
\mathbb{R}
^{n}$; 2) For all $c\in \mathcal{C}_{\mathcal{M}}$, given any one-to-one
mapping $\xi $ from $\mathcal{M}$ to $%
\mathbb{R}
^{n}$, we have that $\xi \in \mathcal{C}_{\mathcal{M}}\Leftrightarrow \xi
\circ c^{-1}$ is a $%
\mathbb{C}
^{\infty }$ diffeomorphism. In this article, the points of $\mathcal{M}$ are
probability distributions. Furthermore, we consider Riemannian manifolds $%
\left( \mathcal{M}\text{, }g\right) $. The Riemannian metric $g$ is not
naturally determined by the structure of $\mathcal{M}$ as a manifold. In
principle, it is possible to consider an infinite number of Riemannian
metrics on $\mathcal{M}$. A fundamental assumption in the information
geometric framework is the choice of the Fisher-Rao information metric as
the metric that underlies the Riemannian geometry of probability
distributions \cite{amari, fisher, rao}, namely%
\begin{equation}
g_{\mu \nu }\left( \Theta \right) \overset{\text{def}}{=}\int dXp\left(
X|\Theta \right) \partial _{\mu }\log p\left( X|\Theta \right) \partial
_{\nu }\log p\left( X|\Theta \right) =-\left( \frac{\partial ^{2}\mathcal{S}%
\left( \Theta ^{\prime }\text{, }\Theta \right) }{\partial \Theta ^{\prime
\mu }\partial \Theta ^{\prime \nu }}\right) _{\Theta ^{\prime }=\Theta }%
\text{,}  \label{FR}
\end{equation}%
with $\mu $, $\nu =1$,..., $n$ for an $n$-dimensional manifold; $\partial
_{\mu }=\frac{\partial }{\partial \Theta ^{\mu }}$ and $\mathcal{S}\left(
\Theta ^{\prime }\text{, }\Theta \right) $ represents the logarithmic
relative entropy \cite{ARIEL},%
\begin{equation}
\mathcal{S}\left( \Theta ^{\prime }\text{, }\Theta \right) =-\int dXp\left(
X|\Theta ^{\prime }\right) \log \left( \frac{p\left( X|\Theta ^{\prime
}\right) }{p\left( X|\Theta \right) }\right) \text{.}
\end{equation}%
The quantity $X$ labels the microstates of the system. The choice of the
information metric can be motivated in several ways, the strongest of which
is Cencov's characterization theorem \cite{cencov}. In this theorem, Cencov
proves that the information metric is the only Riemannian metric (except for
a constant scale factor) that is invariant under a family of
probabilistically meaningful mappings termed congruent embeddings by Markov
morphism \cite{cencov, campbell}.

A geodesic on a $n$-dimensional curved statistical manifold $\mathcal{M}_{S}$
represents the maximum probability path a complex dynamical system explores
in its evolution between initial and final macrostates $\Theta _{\text{%
initial}}$ and $\Theta _{\text{final}}$, respectively. Each point of the
geodesic represents a macrostate parametrized by the macroscopic dynamical
variables $\Theta \equiv \left( \theta ^{1}\text{,..., }\theta ^{n}\right) $
defining the macrostate of the system. Each component $\theta ^{k}$ with $%
k=1 $,..., $n$ is solution of the geodesic equation, $\frac{d^{2}\theta ^{k}%
}{d\tau ^{2}}+\Gamma _{lm}^{k}\frac{d\theta ^{l}}{d\tau }\frac{d\theta ^{m}}{%
d\tau }=0$. Furthermore, each macrostate $\Theta $ is in a one-to-one
correspondence with the probability distribution $p\left( X|\Theta \right) $%
. This is a distribution of the microstates $X$. The set of macrostates
forms the parameter space $\mathcal{D}_{\Theta }$ while the set of
probability distributions forms the statistical manifold $\mathcal{M}_{S}$.
Applications of the IGAC using the IGE\ as a suitable indicator of temporal
complexity (chaoticity) appear in \cite{carlo-MPLB, carlo-PA, carlo-IJTP,
carlo-PD, carlo-EJTP, carlo-CSF}.

\section{The Information Geometric Entropy}

In this Section we focus on the key-steps leading to the construction of the
IGE and comment on its conceptual interpretation.

\subsection{Preliminaries}

Once the distances among probability distributions have been assigned using
the metric tensor $g_{\mu \nu }\left( \Theta \right) $, a natural next step
is to obtain measures for extended regions in the space of distributions.
Consider an $n$-dimensional volume of the statistical manifold $\mathcal{M}%
_{s}$ of distributions $p\left( X|\Theta \right) $ labelled by parameters $%
\Theta ^{\mu }$ with $\mu =1$,..., $n$. The parameters $\Theta ^{\mu }$ are
coordinates for the point $p$ and in these coordinates it may not be obvious
how to write an expression for a volume element $d\mathcal{V}_{\mathcal{M}%
_{s}}$. However, within a sufficiently small region any curved space looks
flat. That is to say, curved spaces are "locally flat". The idea then is
rather simple: within that very small region, we should use Cartesian
coordinates wherein the metric takes a very simple form, namely the identity
matrix $\delta _{\mu \nu }$. In locally Cartesian coordinates $\chi ^{\alpha
}$ the volume element is given by the product $d\mathcal{V}_{\mathcal{M}%
_{s}}=d\chi ^{1}d\chi ^{2}$.....$d\chi ^{n}$, which in terms of the old
coordinates read,%
\begin{equation}
d\mathcal{V}_{\mathcal{M}_{s}}=\left\vert \frac{\partial \chi }{\partial
\Theta }\right\vert d\Theta ^{1}d\Theta ^{2}\text{ with }d\Theta
^{n}=\left\vert \frac{\partial \chi }{\partial \Theta }\right\vert
d^{n}\Theta \text{.}
\end{equation}%
The problem at hand then is the calculation of the Jacobian $\left\vert 
\frac{\partial \chi }{\partial \Theta }\right\vert $ of the transformation
that takes the metric $g_{\mu \nu }$ into its Euclidean form $\delta _{\mu
\nu }$. Let the new coordinates be defined by $\chi ^{\mu }=\Xi ^{\mu
}\left( \Theta ^{1}\text{,...., }\Theta ^{N}\right) $. A small change $%
d\Theta $ corresponds to a small change $d\chi $,%
\begin{equation}
d\chi ^{\mu }=X_{m}^{\mu }d\Theta ^{m}\text{ where }X_{m}^{\mu }\overset{%
\text{def}}{=}\frac{\partial \chi ^{\mu }}{\partial \Theta ^{m}}
\end{equation}%
and the Jacobian is given by the determinant of the matrix $X_{m}^{\mu }$, $%
\left\vert \frac{\partial \chi }{\partial \Theta }\right\vert =\left\vert
\det \left( X_{m}^{\mu }\right) \right\vert $. The distance between two
neighboring points is the same whether we compute it in terms of the old or
the new coordinates, $dl^{2}=g_{\mu \nu }d\Theta ^{\mu }d\Theta ^{\nu
}=\delta _{\alpha \beta }d\chi ^{\alpha }d\chi ^{\beta }$. Therefore the
relation between the old and the new metric is $g_{\mu \nu }=\delta _{\alpha
\beta }X_{\mu }^{\alpha }X_{\nu }^{\beta }$. Taking the determinant of $%
g_{\mu \nu }$, we obtain $g\overset{\text{def}}{=}\det \left( g_{\mu \nu
}\right) =\left[ \det \left( X_{\mu }^{\alpha }\right) \right] ^{2}$ and
therefore $\left\vert \det \left( X_{\mu }^{\alpha }\right) \right\vert =%
\sqrt{g}$. Finally, we have succeeded in expressing the volume element
totally in terms of the coordinates $\Theta $ and the known metric $g_{\mu
\nu }\left( \Theta \right) $, $dV_{\mathcal{M}_{s}}=\sqrt{g}d^{n}\Theta $.
Thus, the volume of any extended region on the manifold is given by,%
\begin{equation}
\mathcal{V}_{\mathcal{M}_{s}}=\int d\mathcal{V}_{\mathcal{M}_{s}}=\int \sqrt{%
g}d^{n}\Theta \text{.}
\end{equation}%
Observe that $\sqrt{g}d^{n}\Theta $ is a scalar quantity and is therefore
invariant under orientation preserving general coordinate transformations $%
\Theta \rightarrow \Theta ^{\prime }$. The square root of the determinant $%
g\left( \Theta \right) $ of the metric tensor $g_{\mu \nu }\left( \Theta
\right) $ and the flat infinitesimal volume element $d^{n}\Theta $ transform
as,%
\begin{equation}
\sqrt{g\left( \Theta \right) }\overset{\Theta \rightarrow \Theta ^{\prime }}{%
\rightarrow }\left\vert \frac{\partial \Theta ^{\prime }}{\partial \Theta }%
\right\vert \sqrt{g\left( \Theta ^{\prime }\right) }\text{, }d^{n}\Theta 
\overset{\Theta \rightarrow \Theta ^{\prime }}{\rightarrow }\left\vert \frac{%
\partial \Theta }{\partial \Theta ^{\prime }}\right\vert d^{n}\Theta
^{\prime }\text{,}  \label{pre1}
\end{equation}%
respectively. Therefore, it follows that,%
\begin{equation}
\sqrt{g\left( \Theta \right) }d^{n}\Theta \overset{\Theta \rightarrow \Theta
^{\prime }}{\rightarrow }\sqrt{g\left( \Theta ^{\prime }\right) }d^{n}\Theta
^{\prime }\text{.}  \label{pre3}
\end{equation}%
Equation (\ref{pre3}) implies that the infinitesimal statistical volume
element is invariant under general coordinate transformations that preserve
orientation (that is, with positive Jacobian).

\subsection{The Formal Construction}

The elements (or points) $\left\{ p\left( X|\Theta \right) \right\} $ of an $%
n$-dimensional curved statistical manifold $\mathcal{M}_{s}$ are
parametrized using $n$ real valued variables $\left( \theta ^{1}\text{,..., }%
\theta ^{n}\right) $,%
\begin{equation}
\mathcal{M}_{s}\overset{\text{def}}{=}\left\{ p\left( X|\Theta \right)
:\Theta =\left( \theta ^{1}\text{,..., }\theta ^{n}\right) \in \mathcal{D}%
_{\Theta }^{\left( \text{tot}\right) }\right\} \text{.}
\end{equation}%
The set $\mathcal{D}_{\Theta }^{\left( \text{tot}\right) }$ is the entire
parameter space (available to the system) and is a subset of $%
\mathbb{R}
^{n}$,%
\begin{equation}
\mathcal{D}_{\Theta }^{\left( \text{tot}\right) }\overset{\text{def}}{=}%
\dbigotimes\limits_{k=1}^{n}\mathcal{I}_{\theta ^{k}}=\left( \mathcal{I}%
_{\theta ^{1}}\otimes \mathcal{I}_{\theta ^{2}}\text{...}\otimes \mathcal{I}%
_{\theta ^{n}}\right) \subseteq 
\mathbb{R}
^{n}
\end{equation}%
where $\mathcal{I}_{\theta ^{k}}$ is a subset of $%
\mathbb{R}
$ and represents the entire range of allowable values for the macrovariable $%
\theta ^{k}$. For example, considering the statistical manifold of
one-dimensional Gaussian probability distributions parametrized in terms of $%
\Theta =\left( \mu \text{, }\sigma \right) $, we obtain%
\begin{equation}
\mathcal{D}_{\Theta }^{\left( \text{tot}\right) }=\mathcal{I}_{\mu }\otimes 
\mathcal{I}_{\sigma }=\left[ \left( -\infty \text{, }+\infty \right) \otimes
\left( 0\text{, }+\infty \right) \right] \subseteq 
\mathbb{R}
^{2}\text{.}
\end{equation}%
In the IGAC, we are interested in a probabilistic description of the
evolution of a given system in terms of its corresponding probability
distribution on $\mathcal{M}_{s}$ which is homeomorphic to $\mathcal{D}%
_{\Theta }^{\left( \text{tot}\right) }$. Assume we are interested in the
evolution from $\tau _{\text{initial}}$ to $\tau _{\text{final}}$. Within
the present probabilistic description, this equivalent to studying the
shortest path (or, in terms of the ME method \cite{c0}, the maximally
probable path) leading from $\Theta \left( \tau _{\text{initial}}\right) $
to $\Theta \left( \tau _{\text{final}}\right) $.

Is there a way to quantify the "complexity" of such path? We propose that
the IGE $\mathcal{S}_{\mathcal{M}_{s}}\left( \tau \right) $ is a good
complexity quantifier \cite{carlo-MPLB, carlo-PA, carlo-IJTP, carlo-PD,
carlo-EJTP, carlo-CSF}. In what follows, we highlight the key-points leading
to the construction of this quantity.

We posit that a suitable indicator of temporal complexity within the IGAC
framework is provided by the \emph{information geometric entropy} (IGE) $%
\mathcal{S}_{\mathcal{M}_{s}}\left( \tau \right) $ \cite{carlo-PD},%
\begin{equation}
\mathcal{S}_{\mathcal{M}_{s}}\left( \tau \right) \overset{\text{def}}{=}\log 
\widetilde{\emph{vol}}\left[ \mathcal{D}_{\Theta }^{\left( \text{geodesic}%
\right) }\left( \tau \right) \right] \text{.}
\end{equation}%
The average dynamical statistical volume $\widetilde{\emph{vol}}\left[ 
\mathcal{D}_{\Theta }^{\left( \text{geodesic}\right) }\left( \tau \right) %
\right] $ is defined as,%
\begin{equation}
\widetilde{\emph{vol}}\left[ \mathcal{D}_{\Theta }^{\left( \text{geodesic}%
\right) }\left( \tau \right) \right] \overset{\text{def}}{=}\lim_{\tau
\rightarrow \infty }\left( \frac{1}{\tau }\int_{0}^{\tau }d\tau ^{\prime }%
\emph{vol}\left[ \mathcal{D}_{\Theta }^{\left( \text{geodesic}\right)
}\left( \tau ^{\prime }\right) \right] \right) \text{,}  \label{rhs}
\end{equation}%
where the "tilde" symbol denotes the operation of temporal average. For the
sake of clarity, we point out that in the RHS of (\ref{rhs}), we intend to
preserve the temporal-dependence by considering the asymptotic leading term
in the limit of $\tau $ approaching infinity. The volume $\emph{vol}\left[ 
\mathcal{D}_{\Theta }^{\left( \text{geodesic}\right) }\left( \tau ^{\prime
}\right) \right] $ is given by,%
\begin{equation}
vol\left[ \mathcal{D}_{\Theta }^{\left( \text{geodesic}\right) }\left( \tau
^{\prime }\right) \right] \overset{\text{def}}{=}\int_{\mathcal{D}_{\Theta
}^{\left( \text{geodesic}\right) }\left( \tau ^{\prime }\right) }\rho
_{\left( \mathcal{M}_{s}\text{, }g\right) }\left( \theta ^{1}\text{,..., }%
\theta ^{n}\right) d^{n}\Theta \text{,}  \label{v}
\end{equation}%
where $\rho _{\left( \mathcal{M}_{s}\text{, }g\right) }\left( \theta ^{1}%
\text{,..., }\theta ^{n}\right) $ is the so-called Fisher density and equals
the square root of the determinant of the metric tensor $g_{\mu \nu }\left(
\Theta \right) $,%
\begin{equation}
\rho _{\left( \mathcal{M}_{s}\text{, }g\right) }\left( \theta ^{1}\text{%
,..., }\theta ^{n}\right) \overset{\text{def}}{=}\sqrt{g\left( \left( \theta
^{1}\text{,..., }\theta ^{n}\right) \right) }\text{.}
\end{equation}%
The integration space $\mathcal{D}_{\Theta }^{\left( \text{geodesic}\right)
}\left( \tau ^{\prime }\right) $ in (\ref{v}) is defined as follows,%
\begin{equation}
\mathcal{D}_{\Theta }^{\left( \text{geodesic}\right) }\left( \tau ^{\prime
}\right) \overset{\text{def}}{=}\left\{ \Theta \equiv \left( \theta ^{1}%
\text{,..., }\theta ^{n}\right) :\theta ^{k}\left( 0\right) \leq \theta
^{k}\leq \theta ^{k}\left( \tau ^{\prime }\right) \right\} \text{,}
\label{is}
\end{equation}%
where $k=1$,.., $n$ and $\theta ^{k}\equiv \theta ^{k}\left( s\right) $ with 
$0\leq s\leq \tau ^{\prime }$ such that,%
\begin{equation}
\frac{d^{2}\theta ^{k}\left( s\right) }{ds^{2}}+\Gamma _{lm}^{k}\frac{%
d\theta ^{l}}{ds}\frac{d\theta ^{m}}{ds}=0\text{.}
\end{equation}%
The integration space $\mathcal{D}_{\Theta }^{\left( \text{geodesic}\right)
}\left( \tau ^{\prime }\right) $ in (\ref{is}) is a $n$-dimensional subspace
of the whole (permitted) parameter space $\mathcal{D}_{\Theta }^{\left( 
\text{tot}\right) }$. The elements of $\mathcal{D}_{\Theta }^{\left( \text{%
geodesic}\right) }\left( \tau ^{\prime }\right) $ are the $n$-dimensional
macrovariables $\left\{ \Theta \right\} $ whose components $\theta ^{k}$ are
bounded by specified limits of integration $\theta ^{k}\left( 0\right) $ and 
$\theta ^{k}\left( \tau ^{\prime }\right) $ with $k=1$,.., $n$. The limits
of integration are obtained via integration of the $n$-dimensional set of
coupled nonlinear second order ordinary differential equations
characterizing the geodesic equations. Formally, the IGE $\mathcal{S}_{%
\mathcal{M}_{s}}\left( \tau \right) $ is defined in terms of a averaged
parametric $\left( n+1\right) $-fold integral ($\tau $ is the parameter)
over the multidimensional geodesic paths connecting $\Theta \left( 0\right) $
to $\Theta \left( \tau \right) $.

In conventional approaches to chaoticity, chaos is specified within the
context of dynamical systems themselves. The existence of classical
dynamical chaos can be inferred from the exponential divergence of the
Jacobi vector field associated to the geodesic flow which coincides with the
natural microscopic dynamics, that is the dynamics described by Newton's
equation of motion. Furthermore, dynamical chaos requires two basic
ingredients: stretching and folding of phase space trajectories. In
geometric language, chaos requires hyperbolicity and compactness of the
manifold where a geodesic flow "lives".

In our information geometric approach to chaos, chaoticity is specified
within the context of suitable statistical manifolds underlying the
probabilistic (entropic) dynamics of dynamical systems when only incomplete
information on the systems is available. Indeed, our approach could have a
wide range of applicability. For instance, as a special limiting case,
Newtonian dynamics can be derived from prior information codified into an
appropriate statistical model \cite{carloAIP}. The basic assumption is that
there is an irreducible uncertainty in the location of particles so that the
state of a particle is defined by a probability distribution. The
corresponding configuration space is a statistical manifold the geometry of
which is defined by the information metric. The trajectory (geodesic)
follows from a principle of inference, the method of Maximum Entropy. No
additional "physical"\ postulates such as an equation of motion, or an
action principle, nor the concepts of momentum and of phase space, not even
the notion of time, need to be postulated.

A geodesic on a curved statistical manifold represents the maximum
probability path a dynamical system explores in its (probabilistic and
statistical) evolution between the initial and the final macrostates on the
statistical manifold. Each point of the geodesic is parametrized by the
macroscopic dynamical variables defining the macrostate of the system.
Furthermore, each macrostate is in a one-to-one relation with the
probability distribution representing the maximally probable description of
the system being considered. The set of macrostates forms the parameter
space while the set of probability distributions forms the statistical
manifold. The parameter space is homeomorphic to the statistical manifold.
The resulting entropic dynamics reproduces the Newtonian dynamics of any
number of particles interacting among themselves and with external fields.
Both the mass of the particles and their interactions are explained as a
consequence of the underlying statistical manifold.

Once again, we point out that several interesting applications of the IGE
appear in the literature \cite{carlo-MPLB, carlo-PA, carlo-IJTP, carlo-PD,
carlo-EJTP, carlo-CSF, carlo-PhD, carluccio}. For instance in \cite%
{carlo-MPLB}, we proposed a novel information-geometric characterization of
chaotic (integrable) energy level statistics of a quantum antiferromagnetic
Ising spin chain in a tilted (transverse) external magnetic field and
conjectured our findings might find some potential physical applications in
quantum energy level statistics. However, in the next Section we will report
our latest conceptual advances in the interpretation of the IGE.

\subsection{The Conceptual Interpretation}

The quantity $\emph{vol}\left[ \mathcal{D}_{\Theta }^{\left( \text{geodesic}%
\right) }\left( \tau ^{\prime }\right) \right] $ is the volume of the
effective parameter space explored by the system at time $\tau ^{\prime }$.
Its faithful geometric visualization may be highly non trivial, especially
in high-dimensional spaces \cite{P02}. The temporal average has been
introduced in order to average out the possibly very complex fine details of
the entropic dynamical description of the system on $\mathcal{M}_{S}$ \cite%
{caves}. Thus, we provide a coarse-grained-like (or randomized-like)
inferential description of the system chaotic dynamics. The long-term
asymptotic temporal behavior is adopted in order to properly characterize
dynamical indicators of chaoticity (for instance, Lyapunov exponents,
entropies, etc.) eliminating the effects of transient effects which enters
the computation of the expected value of $\emph{vol}\left[ \mathcal{D}%
_{\Theta }^{\left( \text{geodesic}\right) }\left( \tau ^{\prime }\right) %
\right] $\textbf{. }In chaotic transients, one observes that typical initial
conditions behave in an apparently chaotic manner for a possibly long time,
but then asymptotically approach a nonchaotic attractor in a rapid fashion.
We term the asymptotic quantity $\widetilde{\emph{vol}}\left[ \mathcal{D}%
_{\Theta }^{\left( \text{geodesic}\right) }\left( \tau \right) \right] $,%
\begin{equation}
\widetilde{\emph{vol}}\left[ \mathcal{D}_{\Theta }^{\left( \text{geodesic}%
\right) }\left( \tau \right) \right] \overset{\text{def}}{=}\lim_{\tau
\rightarrow \infty }\left( \frac{1}{\tau }\int_{0}^{\tau }\left[ \int_{%
\mathcal{D}_{\Theta }^{\left( \text{geodesic}\right) }\left( \tau ^{\prime
}\right) }\rho _{\left( \mathcal{M}_{s}\text{, }g\right) }\left( \theta ^{1}%
\text{,..., }\theta ^{n}\right) d^{n}\vec{\theta}\right] d\tau ^{\prime
}\right) =\exp \left( \mathcal{S}_{\mathcal{M}_{s}}\left( \tau \right)
\right) \text{,}  \label{rhs2}
\end{equation}%
the \emph{information geometric complexity} of the geodesic paths on $%
\mathcal{M}_{S}$. Again, we emphasize that in the RHS of (\ref{rhs2}), we
intend to preserve the temporal-dependence by considering the asymptotic
leading term in the limit of $\tau $ approaching infinity. To have a more
intuitive understanding of $\widetilde{\emph{vol}}\left[ \mathcal{D}_{\Theta
}^{\left( \text{geodesic}\right) }\left( \tau \right) \right] $, we recall
that in going from $\Theta \left( \tau _{0}\right) \overset{\text{def}}{=}%
\Theta _{\text{initial}}$ to $\Theta \left( \tau _{f}\right) \overset{\text{%
def}}{=}\Theta _{\text{final}}$ we assume the system passes through a
continuos succession of (infinitesimally close ) intermediate steps. For
instance at the $\bar{k}$-th step (with $\bar{k}=1$, $2$, $3$,..), we study
the evolution of the system from $\Theta \left( \tau _{\bar{k}-1}\right) $
at $\tau _{\bar{k}-1}$ to $\Theta \left( \tau _{\bar{k}}\right) $ at $\tau _{%
\bar{k}}=$ $\tau _{\bar{k}-1}+d\tau $. At the $\left( \bar{k}+1\right) $-th
step, we study the evolution of the system from $\Theta \left( \tau _{\bar{k}%
}\right) $ at $\tau _{\bar{k}}$ to $\Theta \left( \tau _{\bar{k}+1}\right) $
at $\tau _{\bar{k}+1}=$ $\tau _{\bar{k}}+d\tau $, and so on. Now, let us
consider the following adimensional quantity characterizing two consecutive
steps, the $\bar{k}$-th and $\left( \bar{k}+1\right) $-th steps, 
\begin{equation}
\left( \frac{\delta \widetilde{\emph{vol}}}{\widetilde{\emph{vol}}}\right) _{%
\bar{k}\rightarrow \bar{k}+1}\overset{\text{def}}{=}\frac{\widetilde{\emph{%
vol}}_{\left[ \tau _{\bar{k}\text{, }}\tau _{\bar{k}+1\text{ }}\right] }%
\left[ \mathcal{D}_{\Theta }^{\left( \text{geodesic}\right) }\left( \tau
^{\prime }\right) \right] -\widetilde{\emph{vol}}_{\left[ \tau _{\bar{k}-1%
\text{, }}\tau _{\bar{k}\text{ }}\right] }\left[ \mathcal{D}_{\Theta
}^{\left( \text{geodesic}\right) }\left( \tau ^{\prime }\right) \right] }{%
\widetilde{\emph{vol}}_{\left[ \tau _{\bar{k}-1\text{, }}\tau _{\bar{k}\text{
}}\right] }\left[ \mathcal{D}_{\Theta }^{\left( \text{geodesic}\right)
}\left( \tau ^{\prime }\right) \right] }  \label{rv}
\end{equation}%
where the average infinitesimal statistical volume explored from $\tau _{m}$
to $\tau _{M}$, i.e., $\widetilde{\emph{vol}}_{\left[ \tau _{m\text{, }}\tau
_{M\text{ }}\right] }\left[ \mathcal{D}_{\Theta }^{\left( \text{geodesic}%
\right) }\left( \tau ^{\prime }\right) \right] $, is given by,%
\begin{equation}
\widetilde{\emph{vol}}_{\left[ \tau _{m\text{, }}\tau _{M\text{ }}\right] }%
\left[ \mathcal{D}_{\Theta }^{\left( \text{geodesic}\right) }\left( \tau
^{\prime }\right) \right] \overset{\text{def}}{=}\frac{1}{\tau _{M\text{ }%
}-\tau _{m\text{ }}}\int_{\tau _{m\text{ }}}^{\tau _{M\text{ }}}\emph{vol}%
\left[ \mathcal{D}_{\Theta }^{\left( \text{geodesic}\right) }\left( \tau
^{\prime }\right) \right] d\tau ^{\prime }\text{.}
\end{equation}%
The quantity $\left( \frac{\delta \widetilde{\emph{vol}}}{\widetilde{\emph{%
vol}}}\right) _{\bar{k}\rightarrow \bar{k}+1}$ is the average relative
increment of the volume of the statistical macrospace explored by the system
in its dynamical evolution between two infinitesimally close and consecutive
steps (macrostates). The temporal behavior of $\left( \frac{\delta 
\widetilde{\emph{vol}}}{\widetilde{\emph{vol}}}\right) _{\bar{k}\rightarrow 
\bar{k}+1}$ is a rough indicator of the presence of complex behavior in the
evolution being considered. To have a more reliable complexity indicator, a
step-average over an asymptotically infinite number $N$ of steps would be
required. Therefore, the quantity to consider becomes $\frac{\delta 
\widetilde{\emph{vol}}}{\widetilde{\emph{vol}}}$,%
\begin{equation}
\frac{\delta \widetilde{\emph{vol}}}{\widetilde{\emph{vol}}}\overset{\text{%
def}}{=}\lim_{N\rightarrow \infty }\frac{1}{N}\sum_{k=1}^{N}\left( \frac{%
\delta \widetilde{\emph{vol}}}{\widetilde{\emph{vol}}}\right) _{\bar{k}%
\rightarrow \bar{k}+1}
\end{equation}%
where $\left( \frac{\delta \widetilde{\emph{vol}}}{\widetilde{\emph{vol}}}%
\right) _{\bar{k}\rightarrow \bar{k}+1}$ defined in (\ref{rv}). For
instance, a linear increase of $\frac{\delta \widetilde{\emph{vol}}}{%
\widetilde{\emph{vol}}}$ would be a reasonable manifestation of the presence
of chaoticity (temporal complexity, dynamical stochasticity),%
\begin{equation}
\frac{\delta \widetilde{\emph{vol}}}{\widetilde{\emph{vol}}}\overset{N\text{%
, }\tau \rightarrow \infty }{\approx }\mathcal{K}_{IG}\tau \Leftrightarrow 
\widetilde{\emph{vol}}\overset{N\text{, }\tau \rightarrow \infty }{\approx }%
\exp \left( \mathcal{K}_{IG}\tau \right) \text{.}
\end{equation}%
The quantity $\mathcal{K}_{IG}\overset{\tau \rightarrow \infty }{\approx }%
\frac{d\mathcal{S}_{\mathcal{M}_{s}}\left( \tau \right) }{d\tau }$ is a
model parameter of the complex system and depends on the temporal evolution
of the statistical macrovariables \cite{carlo-MPLB}. It may be interpreted
as playing a role similar to that of the KS entropy rate (sum of all
positive Lyapunov exponents of the dynamical trajectories) and it is, in
principle, an experimentally observable quantity \cite{carlo-MPLB}. We
emphasize there may be physical processes described by several
characteristic time scales where the exponential divergence of $\widetilde{%
\emph{vol}}$ may not be required, although in the presence of chaoticity 
\cite{A96}.

We point out that our construction and interpretation has some similarities
with the logical and thermodynamic depths. For instance, we recall that the
logical depth \cite{B90} is considered to be one of the best candidates as a
measure of (statistical) complexity \cite{L88}. It is an example of a
statistical complexity where the correlated structure of the systems's
constituents play a key role in determining the complex path connecting the
initial and final states of the system under investigation. It is a time
measure of complexity and represents the run time required by a universal
Turing machine executing the minimal program to reproduce a given pattern.
We emphasize that such run time is obtained by a suitable averaging
procedure over the various programs that will accomplish the task by
weighting shorter programs more heavily. Therefore, the logical depth of any
system is defined if a suitably coarse-grained description of it is encoded
into a bit string. In our construction of the IGE, the temporal average has
been introduced in order to average out the possibly very complex fine
details of the entropic dynamical description of the system on $\mathcal{M}%
_{S}$. Therefore, we also provide a coarse-grained-like inferential
description of the system's chaotic dynamics. Furthermore, we point out that
one key objection to the thermodynamic depth \cite{S88} cannot emerge in our
construction of complexity measure. The thermodynamic depth of a process is
a structural measure of complexity and it represents the difference between
the system's coarse- and fine-grained entropy. The "depth" of a macrostate
reached by a particular trajectory is $-const\log p_{j}$ (here $const=k_{B}$
is chosen to be Boltzmann's constant for systems whose successive
configurations can be described in the physical space of statistical
mechanics), where $p_{j}$ is the probability of the $j$-th trajectory. For
the whole range of possible trajectories, the resulting weighted average is $%
-const\sum_{j}p_{j}\log p_{j}$. The set $\left\{ p_{j}\right\} $ represent
probabilities which are consistent with all the measurements that have been
made on the system during its history. This way of reasoning seems very
close in spirit to our complexity measure construction. However, the
key-objection to the thermodynamic depth is the arbitrariness and lack of
explanation of how the macrostates of the system leading to the formation of
the path-trajectory are selected \cite{cosma99}. Instead, in our
construction the selection is as objective as possible since it relies on
the universal ME updating method \cite{c0} where we maximize the logarithmic
relative entropy $\mathcal{S}\left( \Theta _{\bar{k}-1}\text{, }\Theta _{%
\bar{k}}\right) $ \cite{ARIEL} between each pair $\left( \Theta _{\bar{k}-1}%
\text{, }\Theta _{\bar{k}}\right) $ of consecutive macrostates forming the
path connecting the initial $\Theta _{\text{initial}}$ to the final $\Theta
_{\text{final}}$ macrostate. The ME method of determining macroscopic paths
makes no mention of randomness or other incalculable quantities. It simply
chooses the distribution (macrostate) with the maximum entropy allowed by
the information constraints. Thus, it selects the most uninformative
distribution of microstates possible. If we chose a probability distribution
with lower entropy then we would assume information we do not possess; to
choose one with a higher entropy would violate the constraints of the
information we do possess. Thus the maximum entropy distribution is the only
reasonable distribution. Are there other methods of updating? Yes, but the
ME method is the most fundamental, following the rules of probability theory
as outlined by Cox. He proved that probability theory is the only logically
consistent theory of inductive inference \cite{cox}.

\section{Final Remarks}

In this article, we have presented a formal and conceptual reexamination of
the information geometric entropy (IGE), a statistical indicator of temporal
complexity (chaoticity) of dynamical systems in terms of their probabilistic
description on curved statistical manifolds.

In our information geometric approach, the information geometric complexity
represents a statistical measure of complexity of the macroscopic path $%
\Theta \overset{\text{def}}{=}\Theta \left( \tau \right) $ on $\mathcal{M}%
_{S}$ connecting the initial and final macrostates $\Theta _{i}$ and $\Theta
_{f}$, respectively. The path $\Theta \left( \tau \right) $ is obtained via
integration of the geodesic equations on $\mathcal{M}_{S}$ generated by the
universal ME updating method. At a discrete level, the path $\Theta \left(
\tau \right) $ can be described in terms of an infinite continuos sequence
of intermediate macroscopic states, $\Theta \left( \tau \right) =\left[
\Theta _{i}\text{,..., }\Theta _{\bar{k}-1}\text{, }\Theta _{\bar{k}}\text{, 
}\Theta _{\bar{k}+1}\text{,..., }\Theta _{f}\right] $ with $\Theta
_{j}=\Theta \left( \tau _{j}\right) $, determined via the logarithmic
relative entropy maximization procedure subjected to well-specified
normalization and information constraints. The nature of such constraints
define the (correlated) structure of the underlying probability distribution
on the particular manifold $\mathcal{M}_{S}$. In other words, the correlated
structure that may emerge from our information-geometric statistical models
has its origin in the valuable information about the microscopic degrees of
freedom of the actual physical systems. It emerges in the ME maximization
procedure via integration of the geodesic equations on $\mathcal{M}_{S}$ and
is finally quantified in terms of the intuitive notion of volume growth via
the information geometric complexity or, in entropic terms by the IGE. The
information geometric complexity is then interpreted as the volume of the
statistical macrospace explored in the asymptotic limit by the system in its
evolution from $\Theta _{i}$ to $\Theta _{f}$. Otherwise, upon a suitable
normalization procedure that renders the information geometric complexity an
adimensional quantity, it represents the number of accessible macrostates
(with coordinates living in the accessible parameter space) explored by the
system in its evolution from $\Theta _{i}$ to $\Theta _{f}$.

In our view, this work constitutes a non-trivial effort toward an
understanding of the concept of complexity in dynamical systems in terms of
their probabilistic description on curved statistical manifolds through
information geometry and inductive inference. It is also our view that this
effort should be extended to a full quantum domain \cite{carlo-MPLB,
carlo-PA}.

\begin{acknowledgments}
C. C. thanks Ariel Caticha, Akira Inomata, John Kimball, Kevin Knuth, Cosmo
Lupo, Stefano Mancini and Carlos Rodriguez for useful discussions and/or
comments on the IGAC.
\end{acknowledgments}

\end{document}